\documentclass[fleqn, usenatbib]{mnras}

\usepackage{newtxtext,newtxmath}

\usepackage[T1]{fontenc}

\DeclareRobustCommand{\VAN}[3]{#2}
\let\VANthebibliography\thebibliography
\def\thebibliography{\DeclareRobustCommand{\VAN}[3]{##3}\VANthebibliography}

\usepackage{graphicx}	
\usepackage{amsmath}	
\usepackage{soul}

\usepackage{subcaption}

\newcommand{\Msun}{M_\odot}
\newcommand{\Mdot}{\dot{M}}
\newcommand{\Mdotstar}{\dot{M}_\ast}
\newcommand{\Mdotin}{\dot{M}_\mathrm{in}}

\newcommand{\Pdot}{\dot{P}}
\newcommand{\tc}{\tau_{\rm c}}

\newcommand{\Edot}{\dot{E}}

\newcommand{\rin}{r_\mathrm{in}}
\newcommand{\rout}{r_\mathrm{out}}

\newcommand{\rco}{r_\mathrm{co}}
\newcommand{\rinmax}{r_\mathrm{in,max}}
\newcommand{\rA}{r_\mathrm{A}}
\newcommand{\reta}{r_\eta}
\newcommand{\rxi}{r_\xi}

\newcommand{\Rinmax}{R_\mathrm{in,max}}
\newcommand{\DeltaR}{\Delta r/r}

\newcommand{\Lacc}{L_\mathrm{acc}}
\newcommand{\Lcool}{L_\mathrm{cool}}
\newcommand{\Lx}{L_\mathrm{X}}

\newcommand{\Firr}{F_\mathrm{irr}}
\newcommand{\Tp}{T_\mathrm{p}}

\newcommand{\Teff}{T_\mathrm{eff}}
\newcommand{\cs}{c_{\rm s}}
\newcommand{\Md}{M_{\rm d}}

\newcommand{\Gammaacc}{\Gamma_\mathrm{acc}}
\newcommand{\GammaD}{\Gamma_\mathrm{D}}
\newcommand{\Gammadip}{\Gamma_\mathrm{dip}}

\newcommand{\gpers}{g~s$^{-1}$}  
\newcommand{\ergpers}{erg~s$^{-1}$}

\title[Long-term Evolutionary Links]{Long-term Evolutionary Links Between the Isolated Neutron Star Populations}

\author[A. A. Gen\c{c}ali et al.]{
A. A. Gen\c{c}ali$^{1}$\thanks{E-mail: gencali@sabanciuniv.edu}
and \"{U}. Ertan$^{1}$
\\
$^{1}$Sabanc{\i} University, Orhanl{\i}, Tuzla, 34956, \.{I}stanbul, Turkey
}

\date{Accepted XXX. Received YYY; in original form ZZZ}

\pubyear{2024}

\begin{document}
\label{firstpage}
\pagerange{\pageref{firstpage}--\pageref{lastpage}}
\maketitle

\begin{abstract}
We have investigated the evolutionary connections of the isolated neutron star (NS) populations including radio pulsars (RPs), anomalous X-ray pulsars (AXPs), soft gamma repeaters (SGRs), dim isolated NSs (XDINs), ``high-magnetic-field'' RPs (``HBRPs''), central compact objects (CCOs), rotating radio transients (RRATs), and long-period pulsars (LPPs) in the fallback disc model. The model can reproduce these NS families as a natural outcome of different initial conditions (initial period, disc mass, and dipole moment, $\umu$) with a continuous $\umu$ distribution in the $\sim 10^{27} -  5 \times 10^{30}$~G~cm$^3$ range. Results of our simulations can be summarised as follows: (1) A fraction of ``HBRPs'' with relatively high $\umu$ evolve into the persistent AXP/SGR properties, and subsequently become LPPs. (2) Persistent AXP/SGRs do not have evolutionary links with CCOs, XDINs, and RRATs. (3) For a wide range of $\umu$, most RRATs evolve passing through RP or ``HBRP'' properties during their early evolutionary phases. (4) A fraction of RRATs which have the highest estimated birth rate seem to be the progenitors of XDINs. (5) LPPs, whose existence was predicted by the fallback disc model, are the sources evolving in the late stage of evolution before the discs become inactive. These results provide concrete support to the ideas proposing evolutionary connections between the NS families to account for the ``birth-rate problem'', the discrepancy between the cumulative birth rate estimated for these systems and the core-collapse supernova rate.  
\end{abstract}

\begin{keywords}
accretion, accretion discs–methods: numerical–stars: neutron–pulsars: general
\end{keywords}



\section{Introduction}
\label{sec:intro}

The discrepancy between the estimated core collapse supernova (CCSN) rate and the total birth rate of the isolated neutron star (NS) populations, namely normal radio pulsars (RPs), anomalous X-ray pulsars (AXPs), soft gamma repeaters (SGRs), dim isolated NSs (XDINs), ``high-magnetic-field'' RPs (``HBRPs''), central compact objects (CCOs), rotating radio transients (RRATs), and long-period pulsars (LPPs), strongly implies evolutionary connections between some of these NS populations \citep{Faucher2006, Popov2006, Keane2008}. In Fig.~\ref{fig:PPdot_Model2}, these systems are seen in the period-period derivative ($P - \Pdot$) diagram.

AXP/SGRs exhibit sporadic super-Eddington soft gamma-ray bursts and  periodic X-ray pulsations \citep{Cline1980, Hurley1999, Palmer2005, Olausen2014}\footnote{\url{http://www.physics.mcgill.ca/~pulsar/magnetar/main.html}}. For most AXP/SGRs, the rotational power $\Edot = 4 \upi I P \Pdot / P^3$, where $I$ is the moment of inertia of the NS, is lower than the X-ray luminosity, $\Lx$. In the magnetar model \citep{Thompson1995}, it is assumed that AXP/SGRs spin down in vacuum purely by magnetic dipole torques while their X-rays are powered by field decay \citep{Harding2013, Vigano2013, Kaspi2016, Kaspi2017,Enoto2019}. From the dipole-torque formula, the magnetic dipole-field strength at the pole of the star can be estimated as $B_0 = 6.4 \times 10^{19} \sqrt{P \Pdot}$, which gives $B_0 > 10^{14}$~G for most AXP/SGRs. If there is a fallback disc around the NS formed by the supernova matter \citep{Colgate1971, Chevalier1989, Woosley2002, Zeldovich1972}, the dipole torque formula could significantly overestimate $B_0$. The disc formation around the NS is possible if the fallback matter have sufficient angular momentum \citep{Michel1988, Michel1981, Mineshige1997, Perna2014}. In the fallback disc model \citep{Chatterjee2000, Alpar2001}, the disc torque dominates the dipole torque and the source of $\Lx$ is either accretion on to the NS or intrinsic cooling of the NS, depending on the evolutionary phase of the system. For most persistent AXP/SGRs, $\Lx \sim 10^{34} - 10^{36}$~\ergpers~is estimated to be produced by accretion on to the NS. There are also transient AXP/SGRs with much lower $\Lx$ levels in their quiescent states. AXP/SGRs do not show ordinary radio pulsar behaviour. Some AXP/SGRs exhibit transient radio epochs during which they emit radio emissions different from those of ordinary radio pulsars \citep{Camilo2006, Camilo2007_J1810, Camilo2007_1E1547, Kaspi2017}.

So far, seven XDINs have been identified within a distance of $500$~pc. They emit thermal X-rays with $\Lx \sim 10^{30} - 10^{32}$~\ergpers~and black-body temperatures $\Teff = 50 - 100$~eV \citep{Schwope1999, Burwitz2003, Kaplan2003, Haberl+2004, Haberl2007, Rea2007, Turolla2009}. For XDINs, characteristic ages, $\tc \sim {\rm a~few}~10^6$~yr, are greater than their estimated cooling and kinematic ages ($\sim 3 \times 10^{5} - \sim 10^6$~yr) \citep{Motch2009, Tetzlaff2010, Mignani2013, Tetzlaff2011, Tetzlaff2012}. No radio pulsations have been detected from XDINs \citep[for a review see e.g.][]{Turolla2009}. Several XDIN candidates that have no $P$ and $\Pdot$ measurements \citep{Haberl2007, Pires2022, Rigoselli2022, Demasi2024, Kurpas2024, Kurpas2024b} are not included in our analysis.
 
Some RPs are named ``high-magnetic-field'' RPs on the basis of their $B_0$ values deduced from the dipole torque formula alone \citep{Kaspi2017}, are similar to those of AXP/SGRs \citep[$10^{13-15}$~G;][]{Gavriil2008}. In the fallback disc model, their observed properties including estimated braking indices \citep[$n \sim 1 - 3$;][]{Kaspi1994, Camilo2000, Weltevrede2011, Espinoza2011, Antonopoulou2015, Archibald2016} have been obtained with conventional fields \citep{Benli2017, Benli2018}, indicating that ``HBRPs'' are likely to be RPs with relatively high $B_0$ values. They have low $\Lx$ ($< 10^{33}$~\ergpers) and exhibit highly variable  X-ray pulse profiles \citep[e.g.][]{Parent2011, Hu2017}. ``HBRPs'' also display transient behaviour, including outbursts, glitches, and flares \citep{Kaspi2005, Kuiper2009, Parent2011,  Livingstone2011, Ng2012, Antonopoulou2015, Gogus2016, Archibald2016, Archibald2017, Dai2018}. 

CCOs are found at the centres of supernova remnants (SNRs). 
Currently, there are $\sim 10$ known CCOs, and only three of them have measured $P$ and $\Pdot$ values \citep{DeLuca2017}\footnote{\url{https://www.iasf-milano.inaf.it/~deluca/cco/main.htm}}. CCOs display pulsed thermal X-ray emission with $\Lx \sim 10^{33}$~\ergpers~\citep{DeLuca2004, Halpern2010, Gotthelf2013, DeLuca2017}. These sources are known for their persistent surface thermal X-ray emission, and the absence of surrounding pulsar wind nebula \citep{Pavlov2001, DeLuca2017}. CCOs have not been detected in the optical, infrared, and radio bands. 

RRATs emit sporadic radio bursts  with durations in the $0.5 - 100$~ms range and time separations varying from minutes to hours. The  physical mechanism producing  the radio bursts is not known \citep{Mc2006}. The $\Pdot$ values were estimated for $\sim 40$ sources out of more than $100$ known RRATs \citep[ATNF Pulsar Catalogue, version 1.70,][]{Manchester2005}\footnote{\url{https://www.atnf.csiro.au/research/pulsar/psrcat/}}. Only one RRAT was detected in X-rays with  $\Lx \sim 4 \times 10^{33}$~\ergpers~\citep[PSR J1819-1458;][]{Mc2006, Rea2009}. There are $\Lx$ upper limits for several  RRATs  \citep{ Rea2008, Kaplan2009, Keane2013}. The limited  $\Lx$ detections are due to the difficulty in precisely determining their positions \citep{Kaplan2009}.

Recently, four LPPs were discovered with periods $76$~s, $1091$~s, $1318$~s, and $3225$~s \citep{Caleb+2022, Caleb2024, Hurley-Walker2022, Hurley-Walker2023}. They exhibit transient radio-pulsar epochs with unusual radio emission properties \citep{Caleb+2022, Caleb2024, Hurley-Walker2022, Hurley-Walker2023}. Among these sources, $\Pdot$ was estimated only for PSR J0901–4046 \citep{Caleb+2022}. There are $\Pdot$ upper limits for GLEAM-X J162759.5–523504.3 (hereafter GLEAM-X), GPM J1839–10, and ASKAP J193505.1+214841.0 (hereafter ASKAP J1935) \citep{Hurley-Walker2022, Hurley-Walker2023, Caleb2024}. All these sources have only upper limits to $\Lx$.

In our galaxy, the CCSN rate is estimated to be $\beta_{\mathrm{CCSN}}\sim1.9\pm1.1~\mathrm{century}^{-1}$ \citep{Diehl2006, Rozwadowska2021}. Approximately $80\%$ of these events are estimated to form isolated NSs \citep[see][and references therein]{Faucher2006}. The birth rates of different NS populations estimated from statistical calculations and Monte Carlo simulations \citep{Faucher2006, Lorimer2006, Vranesevic2004, Gullon2014, Beniamini2019, Jawor2022} depend on the model assumptions, for the quantities of galactic electron density distribution, the beaming fraction, and active life time of the sources. In most of these works, sources are estimated to be evolving in vacuum with dipole torques. These analyses give the highest birthrate for RRATs ($\beta_{\rm RRATs} = 0.9 - 9.9~\mathrm{century}^{-1}$), followed by RPs ($\beta_{\rm RPs} = 0.9 - 3.3~\mathrm{century}^{-1}$), and XDINs ($\beta_{\rm XDINs} = 2.1 \pm 1.0~\mathrm{century}^{-1}$) \citep{Keane2008, Popov2006, Gill2007, Vranesevic2004, Faucher2006}. The birth rates estimated for the other populations are much smaller \citep[e.g. $\beta_{\rm AXP/SGRs} = 0.3_{-0.2}^{+1.2}~\mathrm{century}^{-1}$;][]{Kouveliotou1998, Gill2007, Ferrario2008, Muno2008, Keane2008}. The sum of the birth rates estimated for the isolated NS populations exceeds the $\beta_{\mathrm{CCSN}}$ \citep{Faucher2006, Popov2006, Keane2008}. It has been proposed that this discrepancy could be due to evolutionary connections, particularly between RPs, RRATs, and XDINs \citep{Faucher2006, Popov2006, Keane2008}. 

The fallback disc model was suggested to explain the $P$ clustering and the $\Lx$ levels of AXPs. In this model, reasonable results are obtained with conventional dipole field strengths of young NSs \citep{Chatterjee2000}. \citet{Alpar2001} proposed that the emergence of AXPs, as well as the other isolated NS populations could be explained if the possibilities of fallback discs are added to the initial conditions (the initial period and dipole moment of the NS). The emission properties of fallback discs have been investigated for various sources \citep{Perna2000, Ertan+2006, Ertan2007, Ertan+2017, Posselt2018, Ozsukan2014}. It was demonstrated that broadband emission from the AXP 4U 0142+61 from the optical to mid-infrared wavelengths \citep{Hulleman2000, Hulleman2004, Morii2005, Wang2006, Kaplan2009ApJ} is in good agreement with the emission from a viscously active and X-ray irradiated disc extending from an inner disc radius $\rin \simeq 10^9$~cm to an outer disc radius $\rout > 10^{12}$~cm \citep{Ertan2007}. It was also shown through detailed calculations that the accretion processes close to the poles of the NS could reproduce the observed soft and hard X-ray spectra in AXP/SGRs consistently with their energy-dependent pulse profiles \citep{Trumper2010, Trumper2013, Kylafis2014, Guo2015, Zesas2015}. 

The fallback disc model was further developed by including the effects of X-ray irradiation, inactivation of the disc at low temperatures, and intrinsic cooling of the NS on the long-term evolution of the sources \citep{Ertan2009}. The model was applied to the individual members of all the NS families \citep{Ertan2014, Benli2016, Benli2017, Benli2018, BenliCCO2018, Ozcan2020, Gencali2018, Gencali2021, Gencali2022, Gencali2023}. These results indicated that the differences in the initial conditions yield different evolutionary avenues leading to different isolated NS populations as proposed by \citet{Alpar+2001}, which also hinted at some evolutionary connections between these systems. For instance, the evolution of some AXP/SGRs with relatively strong magnetic fields towards LPP properties was predicted by \citet{Benli2016} before the discovery of the LPPs \citep{ Caleb+2022, Caleb2024, Hurley-Walker2022, Hurley-Walker2023}. It was also seen from the model simulations that there could be evolutionary connections between RRATs and XDINs \citep{Gencali2021}.

In this work, our focus will be on all possible long-term evolutionary paths connecting two or more isolated NS populations. In Section~\ref{sec:model}, we describe the details of the model. We discuss our results in Section~\ref{sec:results} and summarise the conclusions in Section~\ref{sec:conc}.

\section{The model}
\label{sec:model}

Details of the long-term evolution model applied earlier to different NS populations can be found in our earlier work \citep[see e.g.][]{Ertan2014, Benli2016}. In these earlier applications, we used a conventional approach to determine the inner disc radius and made some simplifications in the calculations of the critical conditions for the accretion/propeller transition and torques in the propeller phase. Earlier results were consistent with the observations within the model simplifications. We will call this model ``Model~I''.

In a later work, an analytical model was developed by \citet{Ertan2017} for a more realistic inner disc radius, $\rin$, calculation for the propeller phase. The $\rin$ values estimated with this model is consistent with the ongoing accretion on to the accreting millisecond X-ray pulsars (AMXPs) and transitional millisecond pulsars in the low $\Lx$ regimes \citep{Ertan2017}. Later, the model was developed to include all the rotational phases (strong-propeller, weak propeller, and spin-up phases) and the critical conditions for the transitions between these rotational phases \citep{Ertan2018, Ertan2021}. It was shown that the model can reproduce the torque reversals of 4U 1626-67 \citep{Gencali+2022}, the best source to study the interaction between a geometrically thin accretion disc and the magnetosphere of the NS around the torque reversals without any wind effect \citep[for a detailed discussion see][]{Camero-Arranz2010}. We adopted this analytical model \citep{Ertan2021} in the calculations of our new long-term evolution model (Model~II). We also tested Model~II with the properties of the sources from different populations. Like Model~I, Model~II can also reproduce the individual source properties of NS populations, while Model~II predicts some additional evolutionary connections between the NS systems. Below, we summarise the features of the rotational phases and the critical conditions for the transitions between these phases in Model~II, and the long-term evolution of a NS with a fallback disc.

The solution of the disc diffusion equation is the same as in Model~I \citep[see e.g.][]{Ertan2009}. This solution gives the mass flow rate, $\Mdotin$, at the inner disc \citep[see e.g.][]{Frank2002}. In the diffusion equation, we use the $\alpha$-prescription for the kinematic viscosity, $\nu = \alpha \cs h$, where $\alpha$ is the kinematic viscosity parameter, $\cs$ is the sound speed, and $h$ is the pressure scale height of the disc \citep{Shakura1973}. 

The disc is heated by the internal viscous dissipation and the X-ray irradiation. The X-ray irradiation flux can be written as $\Firr \simeq 1.2 C \Lx /\upi r^2$ where $C$ is the irradiation efficiency parameter \citep{Fukue1992}. The dynamic outer radius, $\rout$, of the active disc is equal to the radius where the effective temperature, $\Teff$, currently equals the critical inactivation temperature,$\Tp$. Starting from the outermost disc, the disc regions become viscously inactive when the local $\Teff$ decreases below $\Tp$. During the long-term evolution, $\rout$ gradually decreases with decreasing $\Firr$. Eventually, the entire disc becomes viscously inactive, the disc-field interaction ceases, and the source evolves with the weak magnetic dipole torques afterwards. 

The total X-ray luminosity $\Lx = \Lcool + \Lacc$, where $\Lcool$ is the the cooling luminosity of the NS,  $\Lacc = G M \Mdotstar / R$ is the accretion luminosity, $G$ is the gravitational constant, $\Mdotstar$ is the mass accretion rate on to the star, $M$ and $R$ are the mass and radius of the NS respectively. In our calculations, we employ the theoretical curves calculated by \citet{Page+2009}. We tested different cooling curves in our simulations. Both Model~I and Model~II can yield the $\Lx$ and rotational properties of the NS families, while the models favour the minimal and maximal cooling curves respectively. The disc parameters, $\alpha$, $\Tp$, and $C$, are likely to be similar for different isolated NS families. In our earlier work, we obtained reasonable results with $\alpha = 0.045$, $\Tp = 50 - 150$~K, and $C = (1-7) \times 10^{-4}$ \citep[see][and refences therein]{Benli2016, Ertan+2017}.

The maximum radius, $\rinmax$, at which this strong propeller mechanism can be sustained is estimated from the relation
\begin{equation}        
        \Rinmax^{25/8}~|1 - \Rinmax^{-3/2}| ~ \simeq ~ 1.26  ~
\alpha_{-1}^{2/5} ~M_{1.4}^{-7/6} ~\Mdot_\mathrm{in,16}^{-7/20}~ \umu_{30} ~ P^{-13/12}
\label{eq:Rin}  
\end{equation}
\vspace{3mm}
\citep{Ertan2017, Ertan2018, Ertan2021}. Here, $\Rinmax = \rinmax / \rco$, $\rco$ is the co-rotation radius, $\alpha_{-1} = \alpha/0.1$, $M_{1.4} = M/1.4 \Msun$, $\Mdot_{{\rm in}, 16} = \Mdotin /10^{16}$~\gpers, and $\umu_{30} = \umu/10^{30}$~G~cm$^3$. 
The actual inner radius is estimated to be $\rin = r_\eta = \eta \rinmax$ where $\eta \lesssim 1$.

The total torque acting on the NS can be written as
\begin{equation}
    \Gamma = \Mdotstar \sqrt{G M \rin} - \frac{\umu^2}{\rin^3} \Bigg(\frac{\Delta r}{\rin} \Bigg) - \frac{2}{3} \frac{\umu^2 \Omega_\ast^3}{c^3}
\label{eq:torque}
\end{equation}
where $c$ is the speed of light, and $\Omega_\ast$ is the angular spin frequency of the NS. Here, the first term is the spin-up torque, $\Gammaacc$, caused by the mass accretion on to the NS. The second term is the spin-down torque, $\GammaD$, resulting from the interaction between the inner disc and the magnetosphere in the inner boundary with radial thickness $\Delta r$. The last term is the magnetic dipole torque $\Gammadip$.  

In Model~II, the system is in the strong propeller (SP) phase when $\reta = \rin > r_1 = 1.26~\rco$. In this phase, magnetic field lines can effectively eject the inflowing matter out of the system along the open field lines. Since there is no mass transfer on the star, $\Mdot_\ast = 0$, $\Gammaacc = 0$, and $\Lx = \Lcool$; and pulsed radio emission is allowed. In this model, the inner disc has a weak dependence on $\Mdotin$ ($\rin \propto \Mdotin^{-14/125}$) in the SP phase. This leads to different torques and $\Pdot$ evolutions which determine the evolutionary links.

When $\rco < \reta \leq r_1$, the field lines cannot throw the matter from the interaction boundary with speeds greater than the escape speed. The matter ejected from the inner boundary falls back to the disc at larger radii. This leads to a pile up, which pushes the inner disc inward until $\rin = \rco$. At this point, the accretion proceeds while the NS continues to spin-down thus entering the weak propeller (WP) phase. In this phase, $\rin = \rco$ and the matter coupling to the magnetic field lines from this radius flows along the field lines to the magnetic poles of the star. Sources could remain in the WP phase for a rather broad range of $\Mdotin$ \citep[for details, see][]{Ertan2021}. In this phase, $\Lx = \Lacc + \Lcool$ and the pulsed radio emission is not allowed. All the torques given in equation (\ref{eq:torque}) are active, and the source slows down in this phase.

For given $P$ and $B_0$, the sources enter the spin-up (SU) phase when $\Mdotin$ increases above a critical level. This happens when the viscous torques dominate the magnetic torque at $\rco$, which is satisfied roughly when $\rxi = \xi \rA \simeq \rco$ ($\xi = 0.5 - 1$). During this transition, $\rin = \rxi = \rco$, and $\rin$ traces $\rxi$ with increasing $\Mdotin$ for a narrow range of $\Mdotin$. We do not go into details of the SU phase here, because isolated NS populations, with few exceptions, are not likely to enter the SU phase during their long-term evolution.

During the WP and SU phases, the pulsed radio emission is expected to be quenched by the mass accretion on to the star. In the SP phase, a given source could emit radio pulses if it has sufficient power. Inside the death valley, depending on $P$, $B_0$, and the magnetic dipole field geometry, each source has its individual death point below which the normal radio pulses are switched off \citep{Chen1993}. For a model source to be a reasonable representation of a particular ``HBRP'', it should achieve observed $\Lx$ and rotational properties simultaneously during the SP phase (no accretion). Furthermore, $B_0$ and current $P$ values should place the source above the pulsar death line.

The differences in the initial conditions ($P_0$, $\Md$, and $B_0$) determine the initial and subsequent phases of evolution. Among these, the rotational evolution of the sources is most sensitive to $B_0$ by means of the torque resulting from the disc-magnetosphere interaction \citep[see e.g.][]{Benli2016, Gencali2021}. For given $B_0$ and $P_0$, the sources with sufficiently high $\Md$ tend to start the evolution in the WP phase. Once a particular source enters the WP phase, it stays in this phase for a long time ($\Delta t \gtrsim 10^5$~yr) and reaching long periods ($P > 1$~s) in most cases. Below a critical $\Mdotin$, accretion stops and the source enters the SP phase. Afterwords, the X-rays are powered by the cooling of the NS. For given $\Md$ and $B_0$, the sources with longer $P_0$ also tend to start the evolution in the WP phase. For given $\Md$ and $P_0$, there are three different evolutionary pathways: (1) Sources with low $B_0$ starts the evolution in the WP phase, and enter the SP phase at a late stage of the evolution. (2) For intermediate $B_0$ values, the sources start the evolution in the SP phase. For these sources, the spin-down torque is not sufficient to take the systems to the WP phase. $\rco$ can never reach $\rin$ while both are increasing in the SP phase. (3) The sources with highest $B_0$ values also start the evolution in the SP phase. Due to the high $B_0$, the disc torque is very efficient to spin-down the NS, such that, $\rco$, that is initially less than $\rin$, increases rapidly and catches $\rin$ at some point of evolution. This switches on the accretion, and takes the system into the WP phase which lasts until the onset of the final SP phase when $\Mdotin$ decreases below a critical level. We now describe the evolutionary connections between different isolated NS families estimated from these evolutionary paths produced by tracing the initial conditions.

\section{Results \& Discussion} 
\label{sec:results}

\begin{figure*}
    \centering
    \includegraphics[width=1\textwidth]{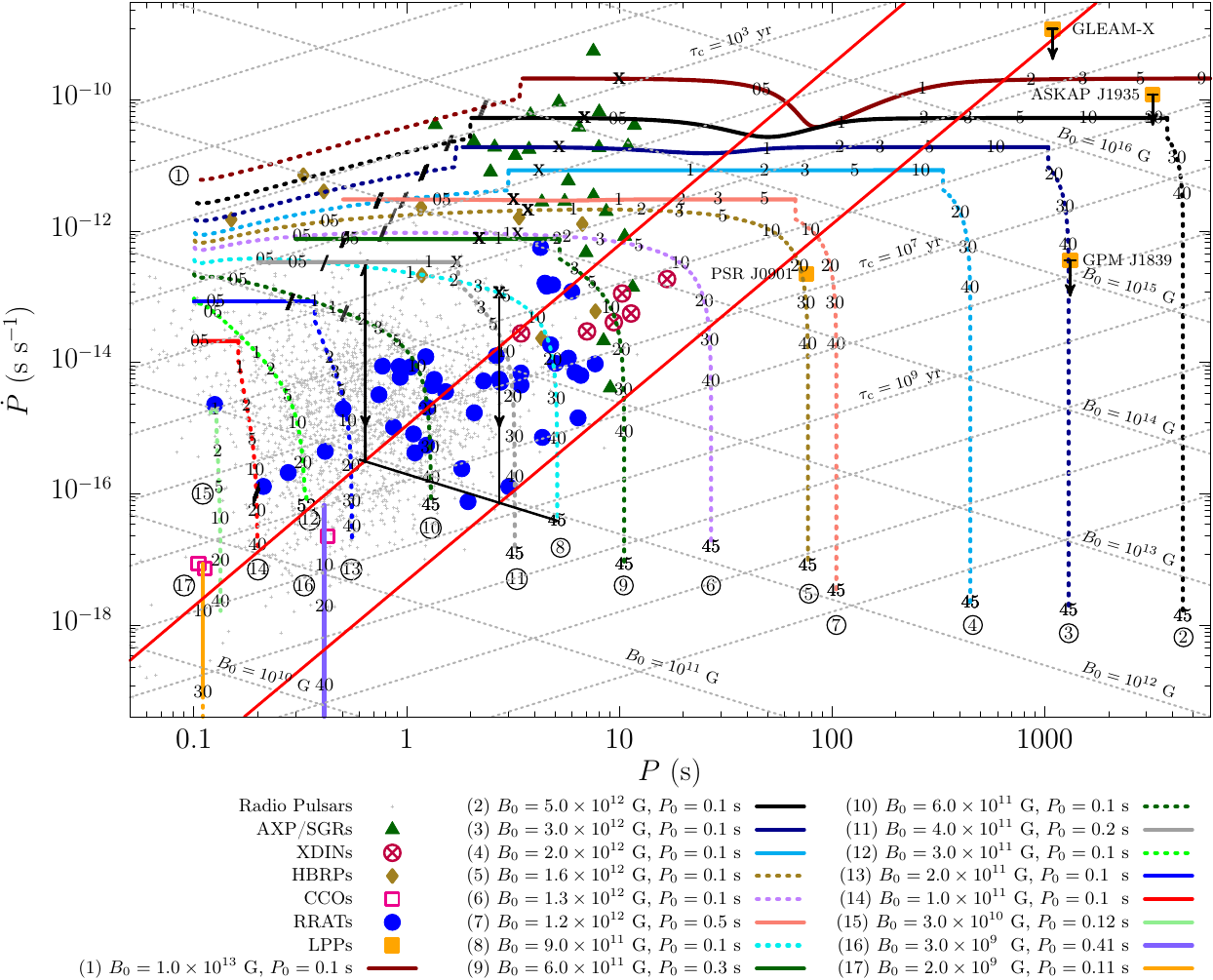}
    \caption{Evolutionary paths in the $P - \Pdot$ diagram. For all models, $\alpha = 0.045$, $C = 1 \times 10^{-4}$, $\Tp = 50$~K, $\DeltaR~=0.25$, $\eta~=0.7~$, and $\xi~=~0.7$. All these curves are generated with $\Md \simeq 1 \times 10^{-5}~M_\odot$ (except for the curves 16 and 17, $\Md \sim 1.6 \times 10^{-6}~M_\odot$). The values of $B_0$ and $P_0$ are given below the figure. The numbers in circles identify the curves for the discussion in the text. Solid and dashed curves correspond to the WP and SP phases respectively. Diagonal, solid (red) lines show the borders of the pulsar death valley \citep{Chen1993}. For comparison, constant $\tau_{\rm c}$ and $B_0$ lines calculated from the dipole-torque formula are also given. The numbers on the model curves show the ages of the sources in units of $10^5$~yr. The markings on the curves indicate the periods at which the sources cross the upper ('/') and lower ('X') boundaries of the death valley in our model. The two solid vertical lines show these points for the model curve $8$. At the end of the evolution, the constant-$B_0$ lines indicate the actual $B_0$ values of the sources. The data for RPs, AXP/SGRs, XDINs, RRATs, ``HBRPs'' and CCOs are obtained from the ATNF Pulsar Catalogue \citep[version 1.70,][]{Manchester2005}. The arrows on the GLEAM-X and GPM J1839 data points show the $\Pdot$ upper limits of the sources.}
    \label{fig:PPdot_Model2}
\end{figure*}

\begin{figure}
    \centering
    \includegraphics[width=1\columnwidth]{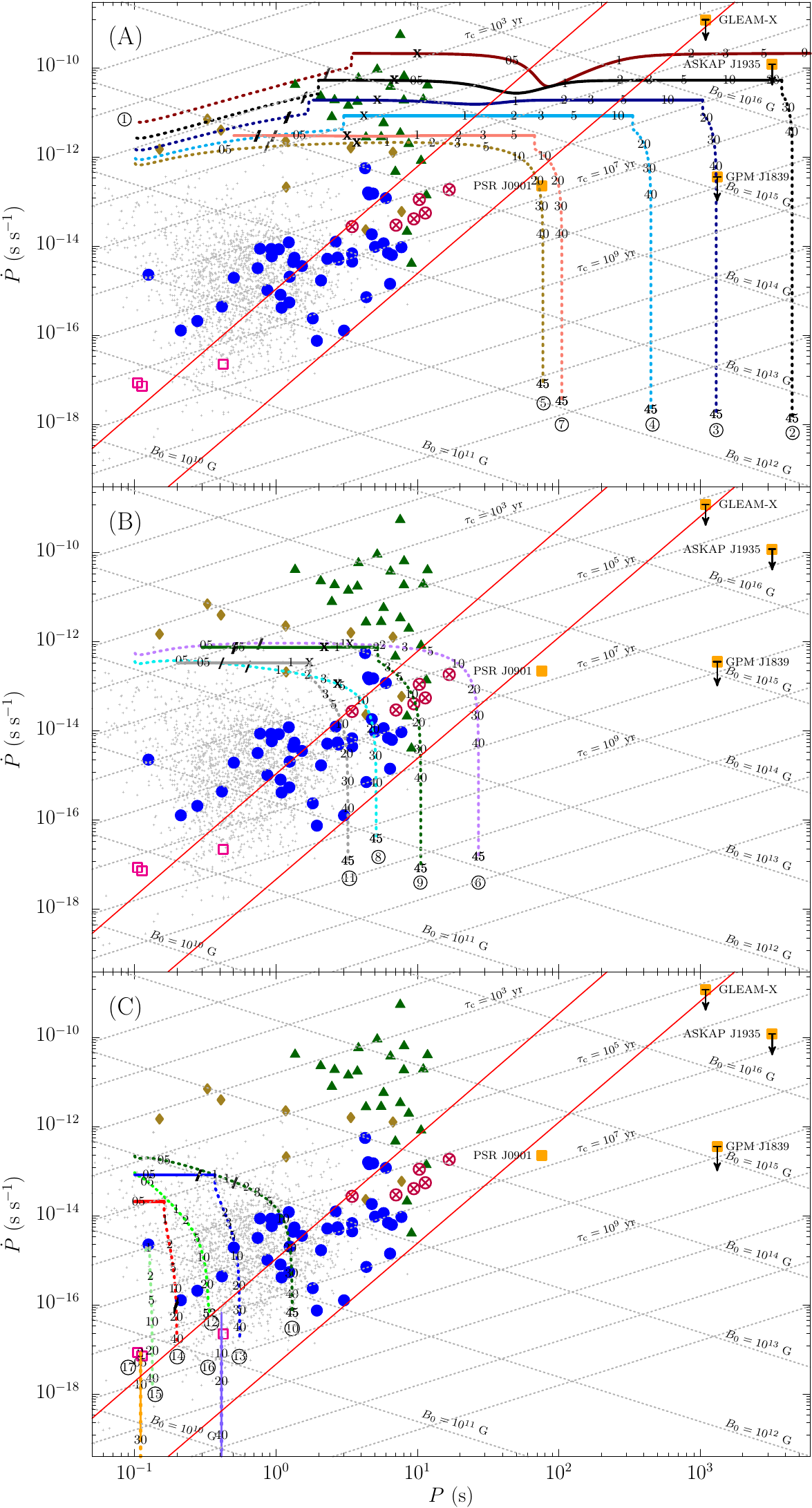}
    \caption{Same as Fig.~\ref{fig:PPdot_Model2}. Each panel show the model curves produced by $B_0$ range: (A) $10^{13}$~G $> B_0 \gtrsim 10^{12}$~G, (B) $10^{12}$~G $\gtrsim B_0 \gtrsim 4 \times 10^{11}$~G, and (C) $6 \times 10^{11}$~G $\gtrsim B_0 \gtrsim 2 \times 10^{9}$~G. See the text for details of the evolutionary connections.}
    \label{fig:PPdot_multi_1x3}
        \parbox[b]{0.03\textwidth}{
        \raisebox{0pt}[0pt][0pt]{\phantomsubcaption\label{fig:2A}}
    }
    \parbox[b]{0.03\textwidth}{
        \raisebox{0pt}[0pt][0pt]{\phantomsubcaption\label{fig:2B}}
    }
    \parbox[b]{0.03\textwidth}{
        \raisebox{0pt}[0pt][0pt]{\phantomsubcaption\label{fig:2C}}
    }
\end{figure}

Figs~\ref{fig:PPdot_Model2} and \ref{fig:PPdot_multi_1x3} show the long-term evolutionary model tracks in the $P - \Pdot$ diagram corresponding to various combinations of initial conditions. All these curves are obtained with the disc parameters $\alpha = 0.045$, $C = 1 \times 10^{-4}$, and $\Tp = 50$~K using the maximal theoretical cooling curve $\Lcool$ estimated by \citet{Page2006, Page+2009}. The solid and dashed curves correspond to the WP phase (accretion allowed) and the SP phase (no accretion) respectively. With a continuous $B_0$ distribution from $\sim 2 \times 10^{9}$~G to $\sim 1 \times 10^{13}$~G, the model yields the rotational characteristics of all the isolated NS populations. All the illustrative model curves in Fig.~\ref{fig:PPdot_Model2} are obtained with the same set of disc parameters given in the figure caption. The differences in the evolutionary paths are due to the different initial conditions, $P_0$ and $B_0$ only. Their values for each model curve are given below the figure. In Fig.~\ref{fig:PPdot_multi_1x3}, Fig.~\ref{fig:PPdot_Model2} is divided into three sub-figures, each corresponding to a different range of $B_0$, which highlights the distinct evolutionary links associated with each $B_0$ range. Below, we summarise these evolutionary links.

\begin{figure}
    \centering
    \includegraphics[width=1\columnwidth]{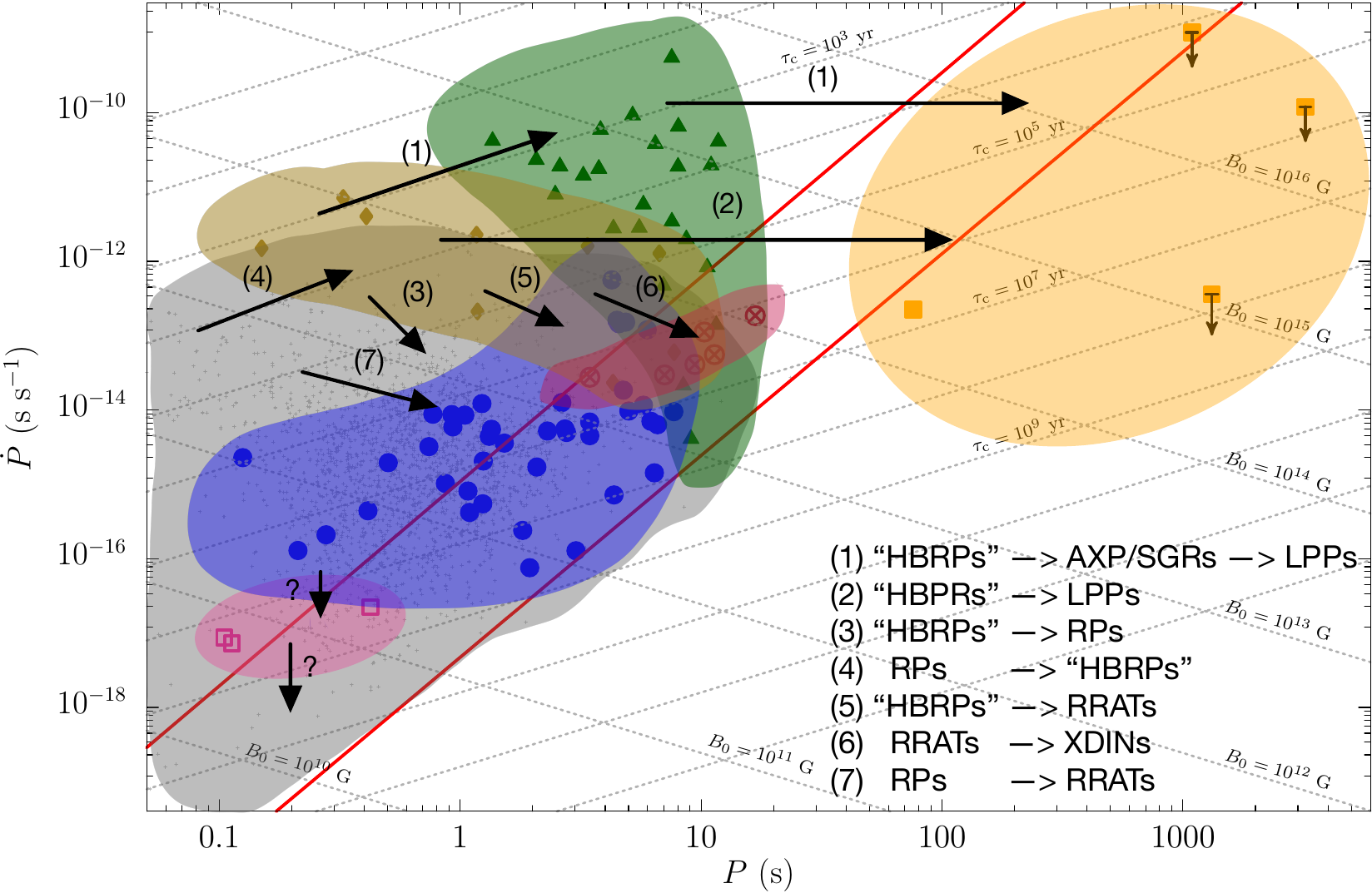}
    \caption{The main evolutionary links between the populations in the $P - \Pdot$ diagram. Shaded areas with different colours roughly indicate the populations with the same colour code used in Fig.~\ref{fig:PPdot_Model2}. The numbered arrows represent possible evolutionary connections between the populations, which are also described in the bottom right corner of the figure. In later phases, CCOs may become RPs and/or RRATs (see the text for the discussion).}
    \label{fig:PPdot_EvolutionaryConnections}
\end{figure}

It is seen that the sources with strongest dipole fields ($1.2 \times 10^{12}$~G~$\sol B_0 \sol 1 \times 10^{13}$~G) pass through the persistent AXP/SGR region (see Figs~\ref{fig:PPdot_Model2} and \ref{fig:2A}; curves $1-5$ and $7$). Most of the known persistent AXP/SGRs seem to evolve in the WP phase (solid curves). In this phase, mass accretion on to the NS prevents ordinary pulsed radio emission. It is seen that a fraction of these sources could be ``HBRPs'' in their young phases of evolution (see e.g. the ``HBRP'' sources on curve 3, with $B_0 = 3 \times 10^{12}$~G). Note that radio pulsations are allowed only during the SP phase (along the dashed curves) below a critical $P$. It is interesting that a particular source starting the evolution in the SP phase as a ``HBRP'' could become an AXP/SGR with the onset of accretion (WP phase), and subsequently spin down significantly to become a LPP like GLEAM-X. For instance, the model curves $1-4$ illustrate this type of evolution. The AXP/SGRs with relatively strong fields could achieve longer $P$ values (see the LPP sources in Fig.~\ref{fig:2A}). Only the sources with $B_0 \sog 3 \times 10^{12}$~G seem to reach $P > 10^3$~s. The LPP sources ASKAP J1935, GLEAM-X, and GPM 1839-10 likely belong to this group (see Fig.~\ref{fig:2A}). It is also seen that a fraction of AXP/SGRs with relatively long $P_0$ and/or high $\Md$ could start the evolution in the WP phase, and never show radio pulsar behaviour over their lifetimes (see curve $7$). Most of these sources enter the final SP phase with long $P$ values placing them below the pulsar death line (lower border of the death valley).

Almost all the sources that have $B_0 \sog 10^{12}$~G and start the evolution in the SP phase are likely to be identified as ``HBRPs'' during the early phases of their evolution (curves $1 - 6$ and $8$). Among these ``HBRPs'', those with relatively high $B_0$ evolve along a path through the persistent AXP/SGR region as explained above (curves $1-4$). The ``HBRPs'' with relatively low $B_0$ may never enter the WP phase, and evolve to the properties of RPs, XDINs, LPPs, and AXP/SGRs with relatively low $\Pdot$ values (curves $5$, $6$, and $8$). These sources cross the pulsar death line before they reach the AXP/SGR region. It is remarkable that a source with $P \simeq 1$~s classified as a ``HBRP'' due to an inferred $B_0 \simeq 10^{14}$~G from the dipole torque model, has a $B_0 \simeq 1.6 \times 10^{12}$~G in our model (see model curve 5, for instance), and is not a ``HBRP'' at all. Thus we use the term ``HBRP'' with the caveat that these are sources for which high $B_0$ values are inferred if the dipole torque is used, while the actual $B_0$ is not necessarily 'high'. It is seen in Fig.~\ref{fig:2A} that the evolutionary curves do not connect persistent AXP/SGRs to RRATs, XDINs, and CCOs (see the curves $1-5$ and $7$). In Fig.~\ref{fig:2B}, the AXP/SGRs with lowest $\Pdot$ are transient sources, and only these systems seem to have evolutionary links with RRATs and XDINs (curves $6$ and $9$). Since AXP/SGRs have relatively low birth-rate, their evolutionary links seems to be unimportant for the solution of the birth-rate problem.

The sources with $B_0 \sim 4 \times 10^{11} - 1.3 \times 10^{12}$~G evolve towards XDIN region (see Fig.~\ref{fig:2B}; curves $6$, $8$, $9$, and $11$) starting the evolution either in the SP phase or in the WP phase. During the early epochs of evolution, these sources would pass through RP, ``HBRP'', RRAT, and possibly transient AXP/SGR evolutionary stages before they become XDINs after $t \sim {\rm a~few~} 10^5$~yr (for the known XDINs). Properties of XDINs are achieved in the SP phase while the sources are evolving below the pulsar death line in the $P - \Pdot$ plane. This means that the reason for the lack of radio pulses from XDINs in this model is their insufficient field strengths and/or long periods rather than the beaming geometry. These sources will become too dim to be observed after their $\Lcool$ levels decrease below the detection limits. In Figs.~\ref{fig:PPdot_Model2} and \ref{fig:PPdot_multi_1x3}, the numbers on the curves represent ages in units of $10^5$~yr. These figures show that the rotational properties of XDINs are achieved at ages $\sim 10^6$~yr. At these ages, $\Lcool \sim 5 \times 10^{31}$~\ergpers~which is mostly consistent with the estimated $\Lx$ levels and kinematic ages of XDINs \citep{Motch2009, Tetzlaff2010, Mignani2013, Tetzlaff2011, Tetzlaff2012}. Fig.~\ref{fig:PPdot_EvolutionaryConnections} shows basic evolutionary links between different populations in a simple schematic picture. The regions with different colours represent different populations. These regions overlap and the borders and are not well defined. The arrows with numbers show the evolutionary links and directions between the populations, which are defined at the bottom right of the figure.

Note that the highest birth rate is estimated for RRATs, followed by XDINs and RPs \citep{Faucher2006, Popov2006, Keane2008}, which means that the presence of evolutionary links between these NS families is especially important to account for the birth-rate problem. The $B_0$ range estimated for RRATs ($5 \times 10^{9}$~G~$\sol B_0 \sol 1.3 \times 10^{12}$~G) fills the gap between the $B_0$ ranges estimated for CCOs and XDINs within the same model (curves $6$ and $8 - 15$). RRAT properties are mostly reached in the SP phase. It is seen in Fig.~\ref{fig:2B} that a small fraction of RRATs will evolve to explain the entire XDIN population (curves $6$, $8$, $9$, and $11$). This implies that a fraction of RRATs with relatively strong fields are the progenitors of XDINs. Furthermore, the model curves crossing the known RRATs trace also almost entire region of RPs in the early phases of evolution before the RRAT phase (Figs~\ref{fig:2B} and \ref{fig:2C}; curves $6$, $8$, and $10 - 15$), indicating a strong evolutionary link between the RP and RRAT populations as well.

We obtained the CCO properties with $B_0$ values of a few $10^9$~G (curves $16$ and $17$). These sources have the weakest dipole moments among the isolated NS populations. With such weak fields and young ages, the $P_0$ values are likely to be close to the current $P$ values. The model can reproduce their rotational properties in the WP phase consistently with the estimated SNR ages and $\Lx$ levels. Depending on $B_0$ and $\Md$ values, some CCOs could start the evolution in a short lasting SU phase \citep[see][]{BenliCCO2018}. After the WP/SP transition, these sources could be identified as RPs or RRATs depending on their death points inside the pulsar death valley. Note that the $P$ values of many RRATs with unknown $\Pdot$ are in the same $P$ range as CCOs. There are also CCO candidates without $P$ and $\Pdot$ measurements (see CCO catalogue\footnote{\url{http://www.iasf-milano.inaf.it/~deluca/cco/main.htm}}). These observations hint at the possibility of the evolutionary connection between RRATs and CCOs as well.

What could be the physical mechanism producing RRAT behaviour? Inside the pulsar death valley, as mentioned in Section~\ref{sec:model}, each RP has its own death point, below which it cannot emit ordinary radio pulses \citep{Chen1993}. For the sources evolving with fallback discs, the disc interact with the magnetosphere, opening up the field lines from the light cylinder radius to $\rin$. This is expected to enhance the magnetic flux of the open field lines through the magnetic poles of the star \citep{Parfrey2016, Parfrey2017}. In connection with this effect, variations in $\Mdotin$ might yield occasional burst-like radio pulsations observed in RRATs and possibly in LPPs for the sources evolving close to and below their death points \citep{Gencali2021, Gencali2022, Gencali2023}. We estimate that the radio bursts likely to terminate below a sufficiently low $\Mdotin$ level such that even the occasional enhancement of the open field flux by the inner disc dynamics is not sufficient to trigger the radio bursts. Detection of these sources after the RRAT phase is not easy due to the low $\Lx$ levels and the lack of radio pulses.

In the magnetar model, NSs in these populations are assumed to evolve in vacuum slowing down by the dipole torques alone \citep[see e.g.][]{Vigano2013}. The $B_0$ values inferred from the dipole torque formula ranges from $\sim 10^{10}$~G for CCOs to above $10^{14}$~G for AXP/SGRs and ``HBRPs''. In this model, while most of CCOs are estimated to show pulsed radio emission, no radio pulses have been detected from known CCOs \citep{DeLuca2017}. This might be due to beaming effects, and the radio pulsars with relatively weak fields ($B_0 < 10^{11}$~G) could be descendents of CCOs \citep{Gotthelf2013b, Luo2015}. To check this possibility \citet{Luo2015} analysed 12 candidate RPs, but could not detect CCO like X-ray emission. To account for the progenitors of CCOs in this model, it was proposed that the magnetic dipole fields of these sources were buried by the accretion of the SN matter, and their dipole fields are currently growing due to reemergence of initially buried fields \citep{Luo2015, Gotthelf2013b, Gotthelf2020, Gourgouliatos2020, Ho2021}. Their fields could be growing to the field strengths of normal radio pulsars \citep{Gotthelf2013b, Luo2015}. However, RPs with thermal X-ray emission that could be immediate descendants of CCOs have not been detected so far \citep{Luo2015}. These results might indicate that CCOs are intrinsically radio quiet for the dipole torque models \citep{Gotthelf2013b, Luo2015}. In the fallback disc model, these sources have $B_0 \sim {\rm a~few}~10^{9}$~G, and the reason for the lack of their pulsed radio emission is that they are the sources currently accreting from their fallback discs. Even after the termination of the accretion, they are not likely to emit pulsed radio emission because such weak dipole fields place them below the pulsar death line. This seems to be consistent with the non-detection of RPs with CCO like X-ray emission. 

Sources with strongest fields are estimated to be AXP/SGRs and ``HBRPs'' in both models while the estimated $B_0$ values are one to two orders of magnitude smaller in the fallback disc model. None of the AXP/SGRs show regular radio pulsations, while some of them exhibit transient or varying radio pulsar epochs with properties rather different from those of normal radio pulsars \citep{Camilo2006, Camilo2007_J1810, Camilo2007_1E1547, Kaspi2017}. The $B_0$ values of ``HBRPs'' deduced from the dipole torque formula are greater than $10^{14}$~G, like those of most AXP/SGRs \citep{Kaspi2017}. In the magnetar model, it is not clear why AXP/SGRs do not show regular radio pulses. The $\Lx$ levels estimated for ``HBRPs'' are close to their estimated $\Lcool$, while $\Lx$ is much greater than $\Lcool$ for most persistent AXP/SGRs. The source of $\Lx$ is field decay in the magnetar model \citep[see e.g.][]{Vigano2013}. In the fallback disc model, most AXP/SGRs cannot produce pulsed radio emission due to accretion on to the star. Some of them that are estimated to be in the SP phase cannot emit radio pulses either, since their fields are much weaker than those estimated in the magnetar model and not sufficient to yield radio pulsations. ``HBRPs'' can emit radio pulses due to their relatively short periods while they are evolving in the SP phase with no accretion on to the NS. For some ``HBRPs'', $\Pdot$ is increasing \citep{Espinoza2011}. In the case of purely dipole torque acting on the source, this requires dipole field growth. This increasing $\Pdot$ behaviour is naturally produced with a constant $B_0$ in the fallback disc model for the sources with relatively strong fields as explained in Section \ref{sec:model}.

XDINs do not show radio pulses either. Their dipole field strengths inferred from the dipole torque formula are sufficient for most of them to produce radio pulses. It was suggested that their radio beams are very narrow due to their long $P$ values, which allows detection of only one out of $40$ sources \citep{Kondratiev2009}. Given that the birth rate of XDINs is comparable to RPs \citep{Faucher2006, Popov2006, Keane2008}, it is not easy to explain the non-detection of other XDINs in the radio band for the dipole torque models. In the fallback disc model, all known XDINs are in the SP phase, while all of them are below the pulsar death line due to relatively weak fields estimated in this model \citep{Ertan2014}.

With the field strengths deduced with the purely dipole torque assumption, a large fraction of RRATs lie beyond the upper border of the death valley. Most of these sources are expected to show regular pulsations in this model, while none of them exhibit typical radio pulsar behaviour. In the fallback disc model, the estimated fields are much weaker and place all RRATs either into the death valley or below the pulsar death line. This is compatible with the possibility that all RRATs could be below their death points inside the death valley \citep{Gencali2021}. The interaction between the inner disc and the magnetosphere increases the flux of the open field lines through the poles of the star \citep{Parfrey2016, Parfrey2017}. This might be the mechanism producing radio bursts \citep{Gencali2021}.

To explain the nature of LPPs, several ideas have been proposed, including highly magnetic white dwarfs \citep{Rea2022}, modified pulsar death line in the presence of fallback disc \citep{Tong2023, Tong2023b}, twisted magnetic fields \citep{Cooper2024}, self-lensing effects in pulsar-black hole binaries \citep{Xiao2024}, magneto-rotational evolution with long $P_0$, and accretion from interstellar medium \citep{Afonina2024}.
It has been also proposed that sources with magnetar fields ($B_0 > 10^{14}$~G) evolving with fallback disc could achieve the observed rotational properties of LPPs \citep{Ronchi2022, Fan2024}. In this scenario, due to strong magnetic fields, the periods of the sources evolve rapidly and reach the current periods at times when $\Lcool$ levels are above the estimated $\Lx$ upper limits \citep[see e.g.][]{Ronchi2022}. Therefore, fast cooling has been proposed for these systems. The $P$ and $\Pdot$ values place most LPPs below the pulsar death line. It was suggested that the mechanism generating unusual radio pulsations of RRATs and AXP/SGRs might also be operating in LPPs as well \citep{Hurley-Walker2022, Hurley-Walker2023}. In the fallback disc model, the existence of LPPs was predicted before their discovery. The evolutionary model curves passing through the AXP/SGR region extends to the properties of LPPs before the inactivation of their discs \citep{Benli2016}.

Our results show that the evolutions of all the isolated NS populations can be explained in a single picture as a natural outcome of the differences in their initial conditions ($P_0$, $\Md$, and $B_0$). The evolutionary links estimated, especially between RPs, RRATs, and XDINs, in the model naturally relieves the discrepancy between the total estimated birthrate of these sources and the SN rate.

\section{Conclusions}
\label{sec:conc}

We have investigated the long-term evolutions and the evolutionary links between the isolated NS populations in the fallback disc model. Our results together with earlier work show that the emergence of the isolated NS families with rather different characteristics can be explained in the fallback disc model in terms of connected evolutionary paths through different phases corresponding to different NS families. Depending on the initial conditions ($P_0$, $B_0$, and $\Md$) a particular NS's evolution will pass through some but not all of the various categories of young NS behaviour. Long-term rotational evolution of a source depends most sensitively on the $B_0$ value. With some exceptions, the sources with higher $B_0$ evolve with higher $\Pdot$  and reach longer $P$ at the end of the lifetime of the viscously active disc ($\sim 4 \times 10^6$~yr). It is remarkable that the model reproduces the basic properties of the isolated NS families with a continuous $B_0$ distribution in the range of a few $10^9$~G~$< B_0 < 10^{13}$~G. We find that AXP/SGRs and ``HBRPs'' have the strongest dipole moments ($B_0 \sog 10^{12}$~G) followed by XDINs and RRATs. CCOs have the lowest $B_0$ values ($\sim$ a few $10^9$~G), while the $B_0$ range estimated for RRATs ($5 \times 10^9$~G~$\sol B_0 \sol 1.3 \times 10^{12}$~G), fill the large gap between those of XDINs and CCOs.

The model curves that can represent long-term evolutions of the populations also reveal the evolutionary connections between these systems which can be summarised as follows: (1) The model sources, that have highest $B_0$ ($\sog 10^{12}$~G), passing through the AXP/SGR region tend to become ``HBRPs'' during the early phases ($10^3$~yr~$\sol t \sol 10^5$~yr) and LPPs during the late phases of evolution (after the AXP/SGR phase). (2) A large fraction of AXP/SGRs seem to have no evolutionary links with RRATs, XDINs, and CCOs. (3) Except persistent AXP/SGRs and LPPs, RRATs have evolutionary links with the other NS families. (4) For a wide range of $B_0$ ($\sim 5 \times 10^9 - 10^{12}$~G), evolutionary curves of known RRATs pass through a large fraction of the RPs during the initial phases, while a small fraction of the RRAT curves ( $4 \times 10^{11}$~G~$\sol B_0 \sol 1.3 \times 10^{12}$~G) traces the region of all known XDINs. This implies that a significant fraction of RPs are evolving into the RRAT properties, and that a small fraction of RRATs are progenitors of the XDINs. (5) CCOs seem to have the weakest dipole fields ($B_0 \sim {\rm a~few~} 10^9$~G) that overlap with the lower tail of the $B_0$ distribution of RRATs. Our results indicate that most CCOs could also experience the RRAT phase during the late evolutionary phases. A schematic representation of these evolutionary connections between the populations is given in Fig.~\ref{fig:PPdot_EvolutionaryConnections}.
  
The results of our model calculations provide concrete support, a proof-of-concept to the earlier ideas proposing that the discrepancy between the CCSN rate and the cumulative birth-rate of the NS families could be due to the evolutionary links between the populations with the highest estimated birth-rates. It is not possible to establish these evolutionary connections to address the birth-rate problem if all isolated young NSs evolved with dipole torques alone. As a final remark, we would like to note that SGR bursts require strong fields as explained in the magnetar model. In the earlier work \citep{Eksi2003, Ertan2009}, it was proposed that these fields could be in the small scale quadrupole components. Presence of these fields is compatible with the fallback disc model, since it is the large-scale dipole component that interacts with the inner disc and governs the rotational evolution of the NS.

In future work we are planning to investigate the $B_0$, $P_0$, and $\Md$ distribution of these sources through population synthesis.

\section*{Acknowledgements}

We thank the referee, Adriana Mancini Pires, for very useful comments that have considerably improved our manuscript. We also thank M. Ali Alpar for useful comments on the manuscript. We acknowledge research support from Sabanc{\i} University, and from T\"{U}B\.{I}TAK (The Scientific and Technological Research Council of Turkey) through grant 120F329. 

\section*{Data Availability}

No new data were analysed in support of this paper.



\bibliographystyle{mnras}
\bibliography{example} 








\bsp	
\label{lastpage}
\end{document}